\documentclass[aps,nofootinbib,twocolumn,preprintnumbers,superscriptaddress,nobibnotes,floatfix,longbibliography]{revtex4-2}

\pdfoutput=1
\usepackage{amsmath,amsfonts,amssymb,mathrsfs}
\usepackage{hyperref,cleveref}
\usepackage[utf8]{inputenc}
\usepackage{esint}

\usepackage{graphicx}
\usepackage{color}
\usepackage{ragged2e}
\usepackage{tikz}

\usepackage[loose]{units}

\newcommand{\f}{f_{\mu\nu}}
\newcommand{\g}{g_{\mu\nu}}

\newcommand{\wt}{\widetilde}

\newcommand{\omegade}{\Omega_\mathrm{DE}}
\newcommand{\mfp}{m_\mathrm{FP}}
\newcommand{\zeq}{z_\mathrm{eq}}
\newcommand{\hgr}{H_\mathrm{GR}}
\newcommand{\hefour}{^4 \mathrm{He}}
\newcommand{\eten}{\eta_{10}}

\newcommand{\omegal}{\Omega_\Lambda}
\makeatletter

\def\hhref#1{\href{http://arxiv.org/abs/#1}{arXiv:#1}}
\usepackage{xstring} 
\newcommand{\hhrefq}[1]{\IfSubStr{#1}{:}{\href{http://inspirehep.net/search?ln=en&ln=en&p=#1&of=hb&action_search=Search&sf=&so=d&rm=&rg=25&sc=0}{InSpires:#1}}{\hhref{#1}}}

\def\art{\@ifnextchar[{\eart}{\oart}}
\def\eart[#1]#2#3#4#5#6{{\rm #2}, {\em #3 \bf #4} {\rm (#6) #5} ({\em #1})}
\def\article{\@ifnextchar[{\earticle}{\oarticle}}
\def\oarticle#1#2#3#4#5#6{{\rm #1}, {``#6''}, {\rm #2 #3 (#5) #4}}
\def\earticle[#1]#2#3#4#5#6#7{{\rm #2}, {``#7''}, {\rm #3 #4 (#6) #5}  [\hhrefq{#1}]}
\def\hepart[#1]#2{{\rm #2, \sl#1}}
\def\heparticle[#1]#2#3{#2, { ``#3''} [\hhrefq{#1}]}

\begin{document}

\title{Constraints on bimetric gravity from Big Bang nucleosynthesis}

\author{Marcus Högås}
\email{marcus.hogas@fysik.su.se}
\affiliation{The Oskar Klein Centre, Department of Physics, Stockholm University, SE 106 91, Stockholm, Sweden}

\author{Edvard Mörtsell}
\affiliation{The Oskar Klein Centre, Department of Physics, Stockholm University, SE 106 91, Stockholm, Sweden}

\begin{abstract}
\noindent Bimetric gravity is a ghost-free and observationally viable extension of general relativity, exhibiting both a massless and a massive graviton. The observed abundances of light elements can be used to constrain the expansion history of the Universe at the period of Big Bang nucleosynthesis. Applied to bimetric gravity, we readily obtain constraints on the theory parameters which are complementary to other observational probes. For example, the mixing angle between the two gravitons must satisfy $\theta \lesssim 18^\circ$ in the graviton mass range $\mfp \gtrsim 10^{-16} \, \mathrm{eV}/c^2$, representing a factor of two improvement compared with other cosmological probes.
\end{abstract}

\maketitle

\section{Introduction}
Since the Universe is currently expanding, even accelerating \cite{Riess:1998cb,Perlmutter:1998np}, tracing the expansion history backwards in time, the Universe must have been in a very hot and dense initial state. These conditions are appropriate for nuclear reactions to take place, leading to the production of the light elements, including D (deuterium) and $\hefour$, referred to as Big Bang nucleosynthesis (BBN). The resulting abundances of these light elements depend on the conditions at this period, for example on the expansion rate of the Universe and the number of neutrino species. Hence, given the observed abundances of light elements, BBN can be used to test non-standard cosmologies.

The cosmological concordance model is based on general relativity (GR) and the standard model of particle physics and has been very successful in accounting for (and predicting) cosmological observations, including the abundances of the light elements \cite{Alpher:1948ve}. However, to fit observations, one introduces two elusive ingredients, the cosmological constant ($\Lambda$) and a cold dark matter (CDM) particle, constituting the $\Lambda$CDM model. The cosmological constant must be extremely fine-tuned in order to almost exactly cancel the quantum contributions to the vacuum energy. The CDM particle makes up the main part of the matter content of the Universe, yet there is no direct detection of this particle(s). Moreover, with increasingly precise observations, a tension has appeared recently between the value of the Hubble constant (i.e., the present-day expansion rate) as inferred from small and large distance measurements, see for example \cite{Verde:2019ivm}. These issues, together with the problem (or impossibility) of quantizing GR, suggest that Einstein's theory of general relativity is not the ultimate theory for gravity.

Bimetric gravity is a natural extension of general relativity, introducing a massive spin-2 field in addition to the massless spin-2 field (identified with the metric in GR) \cite{Hassan:2011zd,Hassan:2012wr}. Accordingly, there are two gravitons, one massive and one massless. The presence of the massless graviton allows gravitational wave observations to be satisfied without severely restricting the parameter space of the theory \cite{Max:2017flc}, as opposed to many other modified gravity models. Bimetric gravity is also compatible with cosmological observations like the cosmic microwave background (CMB), type Ia supernovae (SNIa), and baryon acoustic oscillations (BAO) \cite{vonStrauss:2011mq,Akrami:2012vf,Konnig:2013gxa,Luben:2020xll,Dhawan:2017leu,Lindner:2020eez,Caravano:2021aum,Hogas:2021lns} and it exhibits a screening mechanism that restores general relativity results on solar-system scales \cite{Sjors:2011iv,Enander:2013kza,Babichev:2013pfa,Enander:2015kda,Platscher:2018voh,Luben:2018ekw}. Among the interesting properties of the theory are:
\begin{itemize}
	\item Cosmologies where the accelerated expansion is due to the interaction between the massless and massive spin-2 fields. No cosmological constant is needed \cite{Volkov:2011an,vonStrauss:2011mq,Comelli:2011zm,Volkov:2012wp,Volkov:2012zb,Akrami:2012vf,Volkov:2013roa,Konnig:2013gxa,Hogas:2021fmr,Hogas:2021lns}.\\
	\item  In a certain parameter range, the massive graviton can function as a dark matter particle \cite{Aoki:2016zgp,Babichev:2016hir,Babichev:2016bxi}. Part of the need of dark matter in local environments can also be reduced with a scale-dependent gravitational coupling ``constant'' \cite{Enander:2015kda,Platscher:2018voh,Hogas:2021fmr}.\\
	\item The increased spectrum of cosmological solutions opens up the possibility to ease the Hubble tension. Preliminary studies suggest that the tension can be eased at least to a small degree \cite{Mortsell:2018mfj,Hogas:2021lns}.
\end{itemize}
Challenges to the theory include the appearance of a gradient instability at the linear perturbative level around (homogeneous and isotropic) cosmological backgrounds \cite{Comelli:2012db,Khosravi:2012rk,Berg:2012kn,Sakakihara:2012iq,Konnig:2014dna,Comelli:2014bqa,DeFelice:2014nja,Solomon:2014dua,Konnig:2014xva,Lagos:2014lca,Konnig:2015lfa,Aoki:2015xqa,Mortsell:2015exa,Akrami:2015qga,Hogas:2019ywm,Luben:2019yyx}. This means that structure formation cannot be analyzed using linear perturbation theory, instead it is necessary to solve the full non-linear equations of motion \cite{Aoki:2015xqa,Mortsell:2015exa,Hogas:2019ywm}, requiring a stable (well-posed) form of the equations of motion. Partial progress has been made in that direction \cite{Kocic:2018ddp,Kocic:2018yvr,Kocic:2019zdy,Torsello:2019tgc,Kocic:2019gxl,Torsello:2019jdg,Torsello:2019wyp,Kocic:2020pnm}, but a well-posed form of the equations of motion is still unknown.

The bimetric action (Hassan--Rosen action) is,
\begin{align}
	\mathcal{S}_\mathrm{HR} = \int d^4x &\left[ \frac{1}{2\kappa_g} \sqrt{-g} R + \frac{1}{2 \kappa_f} \sqrt{-f} \wt{R} \right. \nonumber\\
	& \left. - \sqrt{-g} \sum_{n=0}^{4} \beta_n e_n(S) + \sqrt{-g} \mathcal{L}_m\right],
\end{align}
where $g$ and $f$ denote the determinants of the two metrics (spin-2 fields) $\g$ and $\f$, $R$ is the Ricci scalar of $\g$ and $\wt{R}$ the Ricci scalar of $\f$, $\kappa_g$ and $\kappa_f$ are the Einstein gravitational constants of the two metrics\footnote{$\kappa_g=8\pi G/c^4$ in SI-units.}, the $\beta$-parameters are five, free, constant parameters with dimension of curvature ($1/\mathsf{L}^2$), $e_n(S)$ are the elementary symmetric polynomials of the square root $S^\mu{}_\nu$, defined by $S^\mu{}_\rho S^\rho{}_\nu = g^{\mu \rho} f_{\rho \nu}$, and $\mathcal{L}_m$ is the matter Lagrangian. The action is constructed to avoid the Boulware--Deser ghost plaguing generic theories of massive gravity \cite{Boulware:1973my,Hassan:2011zd,Hassan:2011ea,Hassan:2018mbl}.

Traditionally, the $\beta$-parameters are used as the free parameters of bimetric gravity. However, the Hassan--Rosen action is invariant under the rescaling $(\f,\kappa_f,\beta_n) \to (\omega \f, \omega \kappa_f , \omega^{-n/2} \beta_n)$ with $\omega = \mathrm{const}$. Thus, the $\beta$-parameters are not observables and cannot be constrained by observations (without first having to specify a choice of scaling). This problem is circumvented by introducing the rescaling-invariant (dimensionless) parameters,
\begin{equation}
	B_n = \kappa_g \beta_n c^n/ H_0^2,
\end{equation}
where $H_0$ is the Hubble constant and $c$ is the conformal factor between the two metrics in the asymptotic future when the metrics are proportional $\f = c^2 \g$ (see below and Ref.~\cite{Hogas:2021fmr}). With the $B$-parameters, we can define the physical parameters $\Theta = (\theta,\mfp,\omegal,\alpha,\beta)$ \cite{Luben:2019yyx,Luben:2020xll,Hogas:2021fmr}, see Appendix~\ref{sec:PhysParam} for details. These parameters have immediate physical interpretations; $\theta$ is the mixing angle between the massive and massless gravitons. When $\theta=0$, the physical metric $\g$ is aligned with the massless graviton and the theory reduces to GR, whereas if $\theta = 90^\circ$, the physical metric is aligned with the massive graviton and the theory reduces to de Rham--Gabadadze--Tolley (dRGT) massive gravity with a fixed reference metric \cite{deRham:2010ik,deRham:2010kj,Hassan:2011hr,Hassan:2011tf,deRham:2014zqa}. $\mfp$ is the mass of the (massive) graviton measured in units of $H_0 \sim 10^{-33} \, \mathrm{eV}/c^2$, $\omegal$ is the effective cosmological constant that bimetric cosmology approaches in the asymptotic future, and $\alpha$ and $\beta$ are parameters which (among other things) determine the screening mechanism, see Ref.~\cite{Hogas:2021fmr} for details. There is a one-to-one relation between the $B$-parameters and the physical parameters, so due to their immediate physical interpretation, we choose to parameterize the theory in terms of the latter.\\

\noindent \emph{Notation.} We use geometrized units where Newton's gravitational constant and the speed of light are set to one. Subscript zero denotes the present-day value of a quantity. Here, $z$ denotes the redshift $1+z=a_0/a$ and $a$ denotes the scale factor.

\section{Bimetric cosmology}
\begin{figure}[t]
	\centering
	\includegraphics[width=1\linewidth]{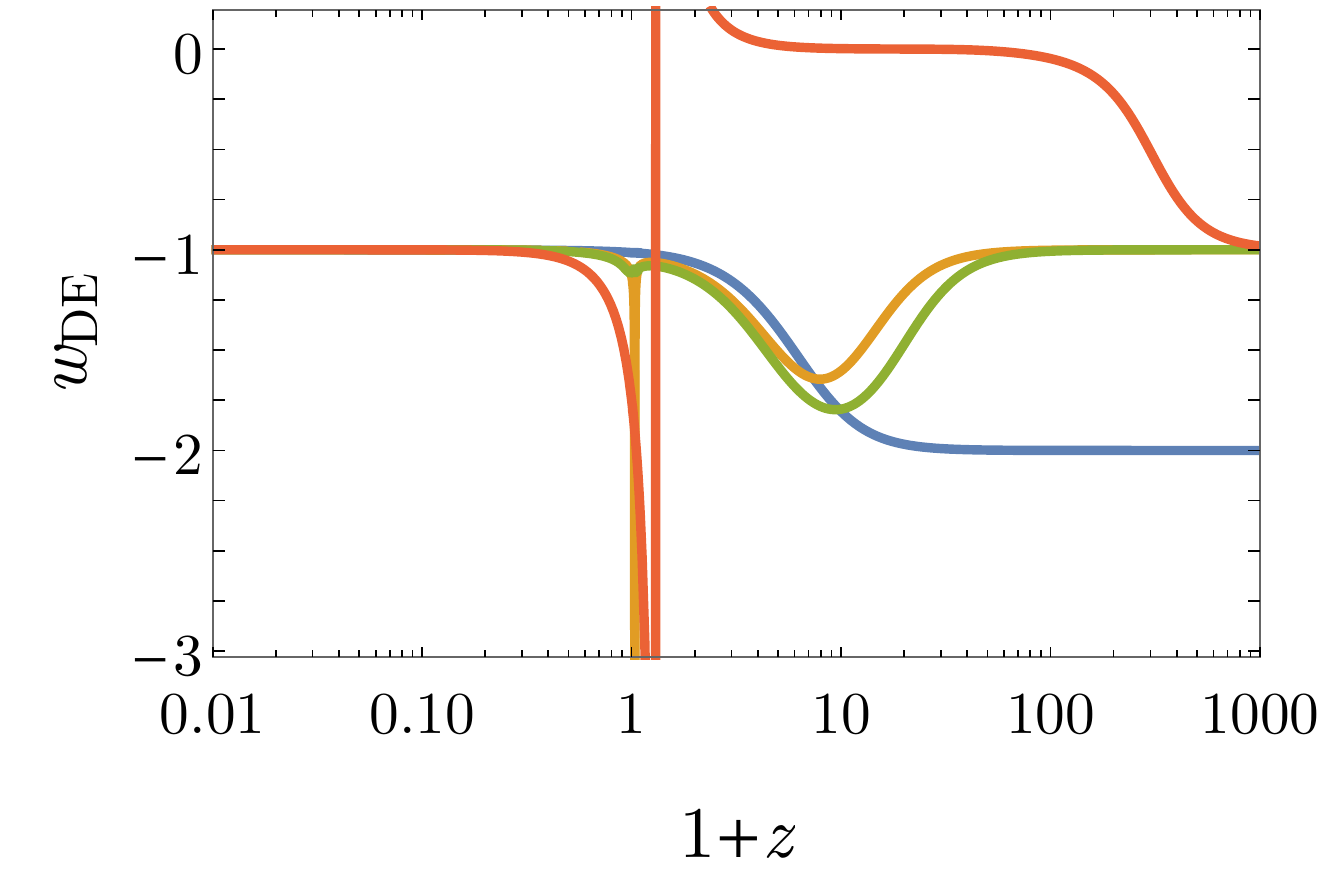}
	\caption{Examples of $w_\mathrm{DE}$ for different physical parameters. Blue curve: $\Theta = (10^\circ,1.2,0.7,1,43)$. Yellow curve: $\Theta = (19^\circ,2.3,0.74,-2.4,8.5)$. Green curve: $\Theta = (19^\circ,2,0.75,-2,8.5)$. Red curve: $\Theta = (45^\circ,10^3,0.7,10,10)$. All models approach $w_\mathrm{DE} = -1$ in the asymptotic future, that is as $1+z\to 0$. The yellow model has a sharp dip at $1+z = 1.03$ reaching a minimum value of $w_\mathrm{DE,min} \simeq -12$. For the red model, $w_\mathrm{DE}$ diverges at $1+z = 1.3$. This is because $\rho_\mathrm{DE}$ changes sign, and hence crosses $\rho_\mathrm{DE}=0$, at this redshift. There is no physical divergence.}
	\label{fig:wde}
\end{figure}

In bimetric gravity, the Friedmann equation is modified by the addition of a dynamical energy density giving a dark energy contribution,
\begin{equation}
	\label{eq:BRFridm}
	\frac{H^2}{H_0^2} = \Omega_m (1+z)^3 + \Omega_r (1+z)^4 + \omegade (1+z)^{3[1+w_\mathrm{DE}(z)]}.
\end{equation}
The equation of state parameter $w_\mathrm{DE}(z)$ depends on time (redshift), with the functional form depending on the physical parameters, see Ref.~\cite{Hogas:2021fmr} for an explicit expression and Fig.~\ref{fig:wde} for some examples. Here, $\Omega_x$ is the present-day energy density of species $x$, measured in terms of the critical density today,
\begin{equation}
	\Omega_x \equiv \rho_{x,0} / \rho_c, \quad \rho_c \equiv 3 H_0^2 /\kappa_g.
\end{equation}
The dark energy density increases with time and contributes to the accelerated expansion of the Universe (hence the subscript ``DE'' for dark energy).

Bimetric cosmology is characterized by a late-universe phase, where the dark energy acts as a cosmological constant with energy density $\omegal$. Thus, the Universe approaches a de Sitter phase in the asymptotic future, as in the $\Lambda$CDM model. In the early universe, $\rho_\mathrm{DE}$ becomes subdominant and the expansion mimics $\Lambda$CDM again. In the intermediate region, $\rho_\mathrm{DE}$ is dynamical. In this phase, there is a wide range of different expansion histories, depending on the physical parameters \cite{Hogas:2021fmr}. The redshift at which this dynamical transition takes place will be referred to as $z_t$ and its value is set by the parameters $(\mfp,\alpha,\beta)$, such that larger values push the transition backwards in time (i.e., to higher redshifts). The redshift at matter-radiation equality is denoted $\zeq$.

If $0 \ll z_t \ll \zeq$ (i.e., $\mfp$, $\alpha$, and $\beta$ not very large), the expansion rate is increased in the region $z_t \lesssim z \ll \zeq$ reaching a maximum value of,
\begin{equation}
	S_\mathrm{max} \simeq 1 + \frac{1}{2}\tan^2 \theta, \quad S \equiv \frac{H}{\hgr},
\end{equation}
see Ref.~\cite{Hogas:2021fmr}. Hence, $S-1$ is the relative difference between the bimetric model and the GR ($\Lambda$CDM) model. Here, we have set the Hubble constant of the two models to be equal. When $z \gtrsim \zeq$, the expansion follows the $\Lambda$CDM model again, implying that there is no modification of the expansion rate at BBN for these values of the parameters.

\begin{figure}[t]
	\centering
	\includegraphics[width=1\linewidth]{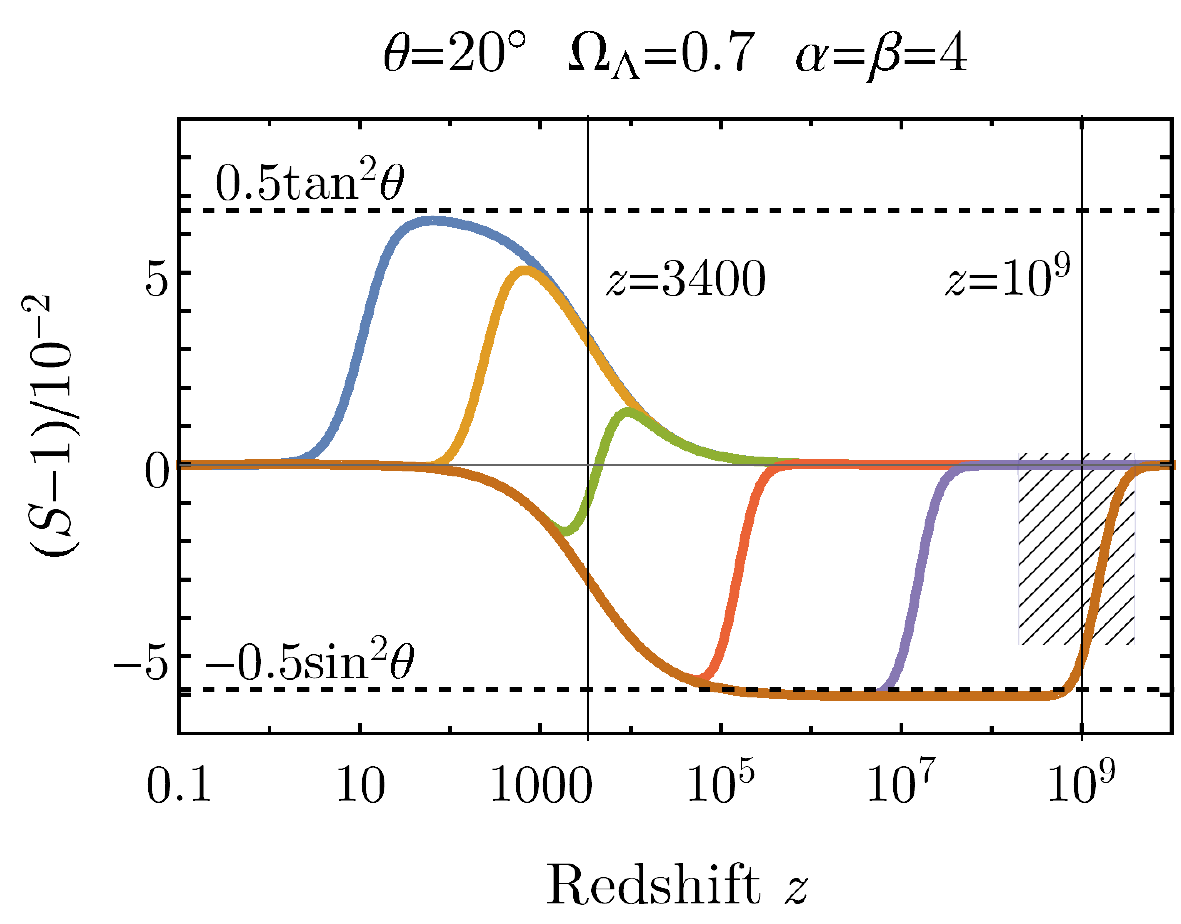}
	\caption{Relative difference (in percent) between the expansion rates of bimetric models and a $\Lambda$CDM concordance model. Increasing values of $\mfp$. Blue: $\mfp = 10$. Yellow: $\mfp = 10^3$. Green: $10^5$. Red: $\mfp = 10^8$. Purple: $\mfp = 10^{12}$. Brown: $\mfp = 10^{16}$. To be compatible with the observational constraint \eqref{eq:Sobs}, the expansion history must lie within the hatched region at BBN ($z \sim 10^9$). The $\mfp = 10^{16}$ model violates these constraints.}
	\label{fig:reldiffh}
\end{figure}

On the other hand, if the transition takes place in the radiation-dominated era $z_t \gg \zeq$ (i.e., if $\mfp$, $\alpha$ or $\beta\gg 1$), the bimetric dark energy mimics a relativistic species with negative energy density in the redshift range $\zeq \lesssim z \lesssim z_t$. Hence, in this regime there is a decreased expansion rate, reaching a minimum value of,
\begin{equation}
	\label{eq:Sapprox}
	S_\mathrm{min} \simeq 1 - \frac{1}{2} \sin^2 \theta.
\end{equation}
See Fig.~\ref{fig:reldiffh} for some examples. When $z \gtrsim z_t$, the dark energy is subdominant and the expansion history follows the $\Lambda$CDM model again. For large enough values of $\mfp$, $\alpha$, or $\beta$, the period of decreased expansion rate extends to the range relevant for BBN (i.e., $z \sim 10^9$), thus setting an upper limit on $\theta$, assuming that $S_\mathrm{BBN} \simeq 1$.

\section{Constraints from BBN}
The baryon-to-photon ratio is parameterized by,
\begin{equation}
	\label{eq:eta10}
	\eten \equiv 10^{10} n_{b} / n_{\gamma},
\end{equation}
where $n_x$ denotes the number density of species $x$. After electron-positron annihilation, the number of baryons and photons in a comoving volume is constant. Hence, the right-hand side of \eqref{eq:eta10} can be evaluated at present day or at BBN, with the same result. The D abundance can be used to determine $\eten$ \cite{Kneller:2004jz}, with the  result \cite{Cyburt:2015mya,Fields:2019pfx},
\begin{equation}
	\label{eq:etenObs}
	\eten = 6.1 \pm 0.2.
\end{equation}
This is consistent with the values derived from CMB data from the Planck satellite \cite{Cyburt:2015mya,Fields:2019pfx}. Whereas D works as a baryometer, $\hefour$ works as a chronometer, restricting the expansion rate, $S$, at BBN. The $\hefour$ abundance can be approximated by the linear relation \cite{Kneller:2004jz,Steigman:2007xt},
\begin{align}
	\label{eq:YpModel}
	Y_p = &0.2386 \pm (0.0006)_\mathrm{num} + \left[100(S-1) + \eten\right]/625,
\end{align}
where the second term on the right-hand side is the numerical error due to the approximation. Here, we assume standard model particle physics with bimetric gravity only modifying the background expansion, $S \neq 1$.

In more recent works, the theoretical model prediction of $Y_p$ is typically parameterized in terms of an effective additional number of neutrino species, $\Delta N_\nu$, rather than in terms of $S$, see for example \cite{Pitrou:2018cgg,Pitrou:2020etk,Cyburt:2015mya,Fields:2019pfx,Yeh:2020mgl,Pisanti:2020efz}. In these works, higher-order corrections to the linear model \eqref{eq:YpModel} are also included. A greater number of neutrino species leads to an increased expansion rate and vice versa for a decreasing number. The modification is achieved by making the replacement $\rho_r \to \rho_r' = \rho_r + \Delta N_\nu \rho_\nu$, which leads to $S^2 = 1 + \Delta N_\nu \rho_\nu / \rho_r$. Before electron-positron annihilation, it reads $S^2 = 1 + 21 \Delta N_\nu / 43$ which can be used to translate between $S \leftrightarrow \Delta N_\nu$.\footnote{Note however, that $S$ is the more fundamental parameter as it allows for a general form of the modified expansion rate.} For our purposes however, the linear model \eqref{eq:YpModel} is sufficient. For example, in the range $S = 0.978 \pm 0.025$ (cf. \eqref{eq:Sobs}), the numerical error of the linear model is only $0.5 \, \%$ compared with the expression including corrections up to third order in $\Delta N_\nu$ \cite{Pitrou:2018cgg}. 

From astrophysical observations, the $\hefour$ abundance is determined to \cite{Aver:2015iza},
\begin{equation}
	\label{eq:YpObs}
	Y_p = 0.2449 \pm 0.0040,
\end{equation}
consistent with the results of Refs.~\cite{Peimbert:2016bdg,Fern_ndez_2018,Fields:2019pfx}. Within GR, $S=1$ and the predicted abundance according to \eqref{eq:YpModel} is $Y_p^\mathrm{GR} = 0.2484 \pm (0.0006)_\mathrm{num}$, slightly higher than, but consistent with, the observed value \eqref{eq:YpObs}. A slower expansion rate ($S<1$) gives more time for neutrons to transform into protons and the lower neutron abundance results in a decrease of the $\hefour$ abundance ($Y_p$). With \eqref{eq:YpModel}-\eqref{eq:YpObs} and using the value \eqref{eq:etenObs}, we derive the following constraint on the expansion rate at BBN,
\begin{equation}
	\label{eq:Sobs}
	S = 0.978 \pm 0.025,
\end{equation}
which can be used to constrain the bimetric parameter space. For models where $H/H_\mathrm{GR}$ is varying during BBN (e.g. the $\mfp = 10^{16}$ model in Fig.~\ref{fig:reldiffh}), $S$ is set to the average value of $H/H_\mathrm{GR}$ within the redshift range $0.2 \times 10^9 \lesssim z \lesssim 3.7 \times 10^9$. Using \eqref{eq:Sapprox}, we predict the constraint $\theta \lesssim 18^\circ$ on the mixing angle in the region where $\mfp$, $\alpha$, or $\beta$ are large enough for $\rho_\mathrm{DE}$ to become subdominant at redshifts greater than BBN ($z \sim 10^9$).

When scanning the parameter space, the present-day matter density ($\Omega_m$) is fixed by the physical parameters (see Ref.~\cite{Hogas:2021fmr}) and the present-day radiation density is set by the current CMB temperature and the neutrino physics, which we assume to be standard, yielding $\Omega_{r} \simeq 9 \times 10^{-5}$. Note however that there is a degree of degeneracy in $Y_p$ between the number of neutrino species and a modified expansion history, as discussed above.

\begin{figure}[t]
	\centering
	\includegraphics[width=\linewidth]{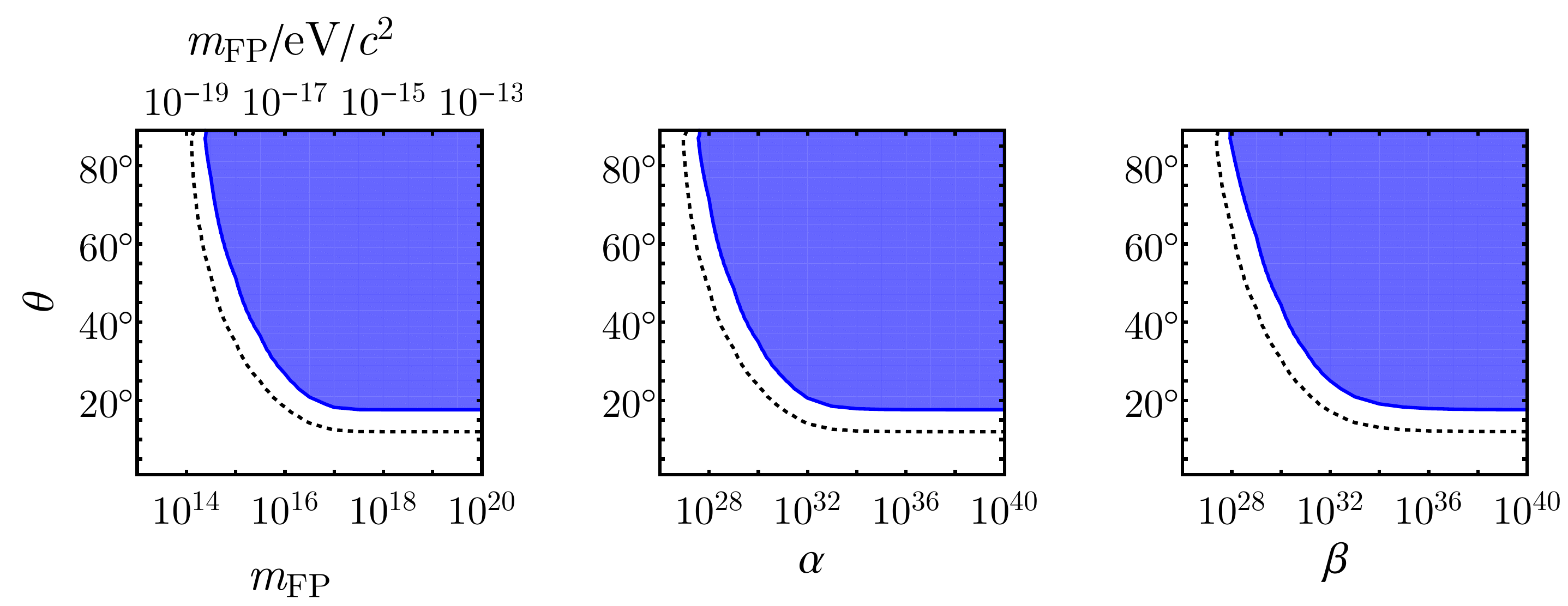}
	\caption{Exclusion plots due to BBN constraints. The dashed curves are the best-fit points where $S=0.978$.}
	\label{fig:exclusionplot}
\end{figure}

The light element abundances are sensitive to the expansion rate in the range $0.2 \times 10^9 \lesssim z \lesssim 3.7 \times 10^9$, corresponding to temperatures $0.06 \, \mathrm{MeV} \lesssim T \lesssim 1 \, \mathrm{MeV}$. We exclude regions in parameter space which violate the observational constraint \eqref{eq:Sobs} in this redshift range. The result is shown in Fig.~\ref{fig:exclusionplot}. It is evident that the GR limit ($\theta \to 0$) is included in the allowed parameter range, but the bimetric model gives a slightly better fit. For example, in the range $\mfp \gtrsim 10^{-16} \, \mathrm{eV}/c^2$, the best fit is $\theta \simeq 12^\circ$, giving a slightly slower expansion rate at BBN compared with GR. When $\mfp \gtrsim 10^{-16} \, \mathrm{eV}/c^2$ or $\alpha \gtrsim 10^{34}$ or $\beta \gtrsim 10^{36}$, the mixing angle must satisfy $\theta \lesssim 18^\circ$, in accordance with our prediction.

\begin{figure}[t]
	\centering
	\includegraphics[width=\linewidth]{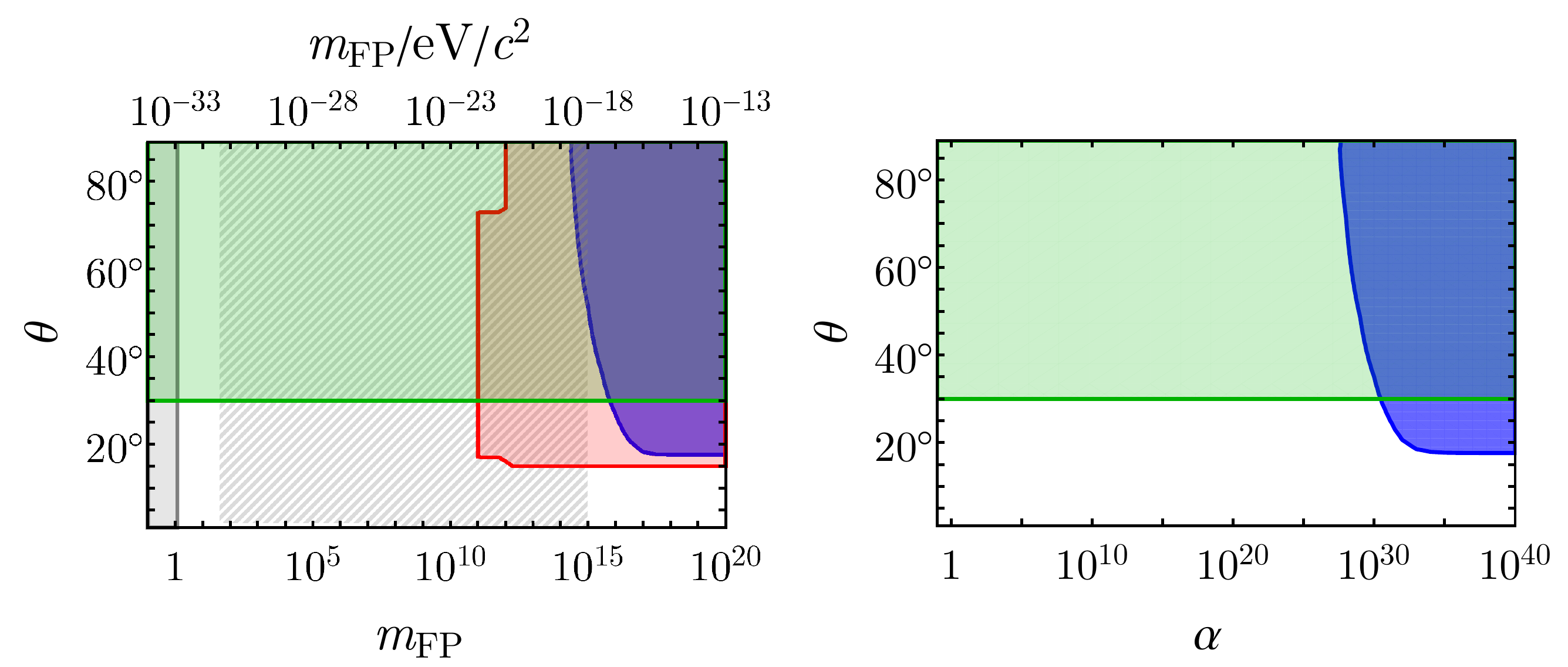}
	\caption{Exclusion plot with compiled constraints from different probes. Blue: BBN. Red: gravitational waves. Green: cosmology (CMB, SNIa, BAO). Gray: the Higuchi bound \cite{Higuchi:1986py}. The hatched region are constraints from local observations including solar-system and galactic tests, assuming $\alpha \sim \beta \sim 1$. If $\alpha$ or $\beta$ are large, these constraints are eliminated. \emph{Left panel}: $\theta \mfp$-plane. \emph{Right panel}: $\theta \alpha$-plane. There is a similar plot in the $\theta \beta$-plane which we do not show here.}
	\label{fig:compiledconstr}
\end{figure}

In Fig.~\ref{fig:compiledconstr}, we compile the BBN results together with constraints from other observational probes. The constraints from gravitational waves (GWs) are from LIGO/Virgo observations of GWs from binary merger systems (neutron star-neutron star or black hole-black hole). The interaction between the massless and massive waves changes the waveform compared with GR. Demanding the waveform to be within observational bounds excludes a graviton mass $10^{11} \lesssim \mfp \lesssim 10^{12}$ in the mixing angle range $17^\circ \lesssim \theta \lesssim 73^\circ$ (cf. \cite{Max:2017flc}). The event GW170817 came with an electromagnetic counterpart \cite{TheLIGOScientific:2017qsa,Monitor:2017mdv}, constraining the propagation speed of gravitational waves to be within one part in $10^{15}$ of the speed of light \cite{GBM:2017lvd}. The observation excludes $\mfp \gtrsim 10^{12}$ if $\theta \gtrsim 70^\circ$. The electromagnetic signal also allows to identify the host galaxy, to which a GW-independent distance measure can be calculated, using the redshift of the galaxy and assuming a cosmological model (in this case $\Lambda$CDM). The distance can also be estimated from the gravitational wave detection. A significant mixing between the massless and massive graviton implies a modified distance. To be compatible with the GW-independent value, we exclude $15^\circ \lesssim \theta \lesssim 75^\circ$ if $\mfp \gtrsim 10^{12}$. The absence of observed gravitational wave echoes can also be used to put constraints on the parameter space. However, since the waveform of the echoes can be very different from the initial GW, it is not obvious that LIGO/Virgo would detect such a signal, rendering the status of these constraints unclear, so we ignore them here.

The cosmological constraint shown in Fig.~\ref{fig:compiledconstr} is based on a conservative fitting of bimetric cosmology to CMB, SNIa, and BAO data \cite{Hogas:2021lns}. If we fix $\alpha$ and $\beta$ to be of order unity, one obtains constraints on the mixing angle and graviton mass from observations in the local universe (e.g., solar-system tests). Here, we take into account solar-system tests, galactic tests, and galaxy cluster lensing and use the results of Refs.~\cite{Luben:2018ekw,Enander:2015kda,Platscher:2018voh}.

From Fig.~\ref{fig:compiledconstr}, it is evident that BBN improves the constraint on the mixing angle by a factor of two (at large $\mfp$) compared with the other cosmological probes with the excluded region from BBN being similar to the constraint from gravitational wave observations. For large values of $\alpha$ or $\beta$ (which are unconstrained by GW observations), BBN improves the existing constraints on the mixing angle by a factor two. The BBN constraints are compatible with the existence of self-accelerating solutions and a screening mechanism (see Ref.~\cite{Hogas:2021fmr} for exact expressions). The allowed parameter region from BBN is also compatible with the massive graviton constituting a dark matter particle \cite{Babichev:2016bxi}.

Interestingly, bimetric gravity is not forced to its GR limits, even after taking all these observational probes into account. For small values of $(\mfp,\alpha,\beta)$, we have $\theta \lesssim 30^\circ$ and for large values of these parameters, we have $\theta \lesssim 15^\circ$, allowing for a significant mixing between the massless and massive gravitons. The best-fit value of $\theta$ from CMB, SNIa, and BAO combined is $\theta = {11^\circ}^{+17^\circ}_{-11^\circ}$ while the best-fit to BBN data is $\theta = {12^\circ}^{+6^\circ}_{-12^\circ}$ (assuming $\mfp \gtrsim 10^{-16} \, \mathrm{eV}/c^2$).

\onecolumngrid

\clearpage

\vspace{5pt}
\subsubsection*{Acknowledgments} 
Thanks to Erik Schildt for many interesting discussions on gravitational waves in bimetric gravity and to an anonymous referee for valuable comments. EM acknowledges support from the Swedish Research Council under Dnr VR 2020-03384.
\vspace{10pt}
\twocolumngrid

%%%%%%%%%%%%%%%%%%%%%%%%%%%%%%%%%%%%%%%%%%%%%%%%%%%%%%%%%
\appendix\section{The physical parameters}
\label{sec:PhysParam}
The physical parameters are expressed in terms of the $B$-parameters as,
\begin{subequations}
	\begin{align}
		\tan^2 \theta &= \frac{B_1 + 3 B_2 + 3 B_3 + B_4}{B_0 + 3 B_1 + 3 B_2 + B_3},\\
		\mfp^2 &= \left(B_1 + 2 B_2+ B_3\right) / \sin^2 \theta,\\
		\omegal &= \frac{B_0}{3} + B_1 + B_2 + \frac{B_3}{3},\\
		\alpha &= - \frac{B_2 + B_3}{B_1 + 2 B_2 + B_3},\\
		\beta &=\frac{B_3}{B_1 + 2 B_2 + B_3}.
	\end{align}
\end{subequations}
The relation can be inverted, resulting in,
\begin{subequations}
	\label{eq:PhysToBs}
	\begin{align}
		B_0 &= 3 \omegal - \sin^2 \theta \, \mfp^2 (3 + 3\alpha + \beta),\\
		B_1 &= \sin^2 \theta \, \mfp^2 (1+ 2 \alpha + \beta),\\
		B_2 &= - \sin^2 \theta \, \mfp^2 (\alpha + \beta),\\
		B_3 &= \sin^2 \theta \, \mfp^2 \beta,\\
		B_4 &= 3 \tan^2 \theta \, \omegal + \sin^2 \theta \, \mfp^2 (-1 + \alpha - \beta).
	\end{align}
\end{subequations}
The Jacobian determinant of \eqref{eq:PhysToBs} is $18 \sin^6 \theta \, \mfp^5 \omegal$, so there is a one-to-one correspondence between the physical parameters and the $B$-parameters, except when $\theta$, $\mfp$, or $\omegal$ are vanishing. 

%%%%%%%%%%%%%%%%%%%%%%%%%%%%%%%%%%%%%%%%%%%%%%%%%%%%%%%%%

%\clearpage

\footnotesize
\bibliographystyle{apsrev}

\bibliography{biblio}
%=========================================================================
%\begin{thebibliography}{nnn}
%\bibitem{Lovelock:1971yv}
%\article[Lovelock:1971yv]{D. Lovelock}{J. Math. Phys.}{12}{498}{1971}
%{The Einstein tensor and its generalizations}.

%\end{thebibliography}

%%%%%%%%%%%%%%%%%%%%%%%%%%%%%%%%%%%%%%%%%%%%%%%%%%%%%%%%
%%%%%%%%%%%%%%%%%%%%%%%%%%%%%%%%%%%%%%%%%%%%%%%%%%%%%%%%

\end{document}